# Toward Low Temperature Solid-Source Synthesis of Monolayer MoS₂

Alvin Tang[1,†], Aravindh Kumar[1,†], Marc Jaikissoon[1], Krishna Saraswat[1,2,3], H.-S. Philip Wong[1,3], and Eric Pop[1,2,3*]

[1]Department of Electrical Engineering, Stanford University, Stanford, CA, 94305, USA

[2]Department of Materials Science and Engineering, Stanford University, Stanford, CA, 94305, USA

[3]Precourt Institute for Energy, Stanford University, Stanford, CA, 94305, USA

[†]Authors contributed equally.

*To whom correspondence should be addressed: *epop@stanford.edu*

## ABSTRACT

Two-dimensional (2D) semiconductors have been proposed for heterogeneous integration with existing silicon technology, however their chemical vapor deposition (CVD) growth temperatures are often too high. Here, we demonstrate direct CVD solid source precursor synthesis of continuous monolayer (1L) MoS₂ films at 560°C in 50-minutes, within the 450-to-600°C, 2-hour thermal budget window required for back-end-of-the-line (BEOL) compatibility with modern silicon technology. Transistor measurements reveal on-state current up to ~140 µA/µm at 1 V drain-to-source voltage for 100 nm channel lengths, the highest reported to date for 1L MoS₂ grown below 600°C using solid source precursors. The effective mobility from transfer length method (TLM) test structures is $29 \pm 5$ cm²V⁻¹s⁻¹ at $6.1 \times 10^{12}$ cm⁻² electron density, which is comparable to mobilities reported from films grown at higher temperatures. The results of this work provide a path towards the realization of high quality, thermal-budget-compatible 2D semiconductors for heterogeneous integration with silicon manufacturing.

**KEYWORDS**: 2D materials, transition metal dichalcogenides, MoS₂, molybdenum disulfide, BEOL, back-end-of-the-line, chemical vapor deposition, CVD growth, carrier mobility

Since the era of the first integrated circuits, improving semiconductor device density has continuously translated into benefits for more advanced computing systems.[1] The horizontally stacked gate-all-around (GAA) nanosheet transistor with 3 nm thick Si nanosheets is expected to replace the FinFET structure to continue transistor scaling beyond the 5 nm technology node.[2] However, further gate length scaling or better control over off-state leakage requires thinner body channel materials,[3] making atomically thin two-dimensional (2D) materials extremely appealing for use in next-generation computing devices. While graphene, the first 2D material discovered, is a semimetal with 0 eV band gap,[4] a class of 2D layered materials known as transition metal dichalcogenides (TMDs) have band gaps ranging from semimetallic (0 eV) to semiconducting (~2.5 eV) energies.[5,6] TMDs retain sizable band gaps and carrier mobilities down to monolayer thicknesses[7] less than 1 nm, whereas Si thinned below ~4 nm faces severe mobility degradation issues due to surface roughness fluctuations.[8–10]

Specifically, $MoS_2$ has become one of the most promising TMDs due to its band gap (~2.2 eV for monolayer $MoS_2$[11,12]) and controllable growth of consistent large-area $MoS_2$ films down to monolayer (1L) thicknesses, making $MoS_2$ a viable candidate for large scale semiconductor applications.[13–18] Although moving $MoS_2$ away from tedious and inconsistent mechanical exfoliation techniques has enabled systematic studies for electronics applications,[16,19,20] the quality of synthesized $MoS_2$ has not always been ideal, especially from the standpoint of the device fabrication thermal budget. If $MoS_2$ is to be integrated in the back-end (i.e. after silicon) of modern integrated circuits, the fabrication process of as-grown $MoS_2$ transistors cannot exceed a range of 450-to-600°C (depending on process time) for integration with logic applications.[21–24] (Although some 3-dimensional vertical memory applications can withstand >700°C.[25]) For example, irreversible degradation of certain silicide contacts, inter-layer dielectrics, and diffusion barrier layers have been observed with thermal budgets over 600°C for 2 hours.[21,26,27]

In this work, we report direct chemical vapor deposition (CVD) growth of continuous monolayer $MoS_2$ films using solid source precursors at 560°C in 50-minutes, within the thermal budget of modern integrated circuit processing. Comparing metrics of $I_{on}/I_{off}$ current ratio, drive current, contact resistance, and carrier mobility, transistors made from these as-grown films show performance levels similar to those reported for monolayer $MoS_2$ grown at higher temperatures.[13,16,28–34] This work provides a path towards the realization of high quality, thermal-budget-compatible TMD materials for applications in transistor density scaling, to further improve new advanced computing systems.

## GROWTH AND MATERIAL CHARACTERIZATION

The thermal budget for modern silicon integrated circuit processing dictates that fabrication of BEOL (back-end-of-the-line) devices and materials cannot exceed 550°C for 2 hours,[26,27] and even more stringent temperature budgets are needed if process times are longer. This limitation is imposed by gate work function instability and silicide contact degradation.[27] Certain interlayer low-κ dielectrics (e.g. SiCOH) could require even stricter thermal budgets, below 450°C for 2 hours.[21] (Although SiCOH integrity could be maintained up to 525°C for 2 hours using some BEOL processes.[24]) Thus, in order to incorporate new materials such as $MoS_2$ into stacked, heterogeneous integrated circuits that increase the transistor density in the third dimension (3D),[35] their BEOL thermal budget becomes crucial. However, direct, monolayer (1L) $MoS_2$ growths with good electrical properties have typically been obtained using solid source precursor chemical vapor deposition (CVD) above 650°C.[13,16,28–34] Wafer-scale 1L $MoS_2$ grown using metal-organic CVD (MOCVD)[14,36–41] at or below 500°C appears within the BEOL temperature budget, but the growth times have been >8 hours, which limit throughput, and the average carrier mobilities reported remain lower than the best $MoS_2$ grown at higher temperatures. Efforts to develop atomic layer deposition (ALD) of $MoS_2$ at temperatures as low as 50°C are also underway,[42,43] but the resulting films are amorphous and require post-growth anneals above 800°C to improve their crystallinity.[44–46]

Here, we grew continuous 1L $MoS_2$ films at 560°C in 50-minutes using solid sulfur (S) and molybdenum trioxide ($MoO_3$) precursors with the aid of perylene-3,4,9,10 tetracarboxylic acid tetrapotassium salt (PTAS).[13,15,17,18] PTAS dissolved in water is deposited as droplets around the perimeter of 1.5 × 2 cm chips with thermally grown $SiO_2$ on p$^+$ silicon, which also serve as back-gate for field-effect transistors.[13] The PTAS droplets dry up into "coffee rings" around the perimeter of the substrate before the seeded $SiO_2$/Si chips are placed face-down on top of an alumina ($Al_2O_3$) crucible containing $MoO_3$ powder. Trace amounts of PTAS diffuse from the dried "coffee rings" towards the center of the chip to facilitate $MoS_2$ nucleation during growth. Excess solid sulfur source melted in a quartz boat is placed at an optimal position upstream near the incoming Ar gas flow inlet, all enclosed within a cylindrical quartz tube.

In addition to the PTAS seed layer promoting initial $MoS_2$ nucleation from which grains grow, the trace amounts of carbon from the diffusing seed layer also act as a catalyst, promoting reduction of $MoO_3$ powder into volatile suboxide $MoO_{3-x}$.[47] The reduction of $MoO_3$ into $MoO_{3-x}$ is crucial for incoming S to react with volatile molybdenum (Mo) species in order to grow $MoS_2$. It is important to note that at growth temperatures around 560°C, $MoO_3$ reduction is near the limit of reaction with trace amounts of carbon present.[47] Both a reduced system pressure of 490 torr and incoming volatile sulfur introduced at elevated temperatures also promote the reduction of $MoO_3$ into volatile suboxide $MoO_{3-x}$.[48] When sufficient volatile $MoO_{3-x}$ is matched with an optimal flux of incoming sulfur atoms at the substrate surface, further reaction occurs to grow 1L $MoS_2$ films. As shown in Figures 1a-b, relatively clean 1L $MoS_2$ films can be grown at 560°C in 50-minutes on 1.5 × 2 cm chips. This is the lowest thermal budget that produced clean, sizable (>10 μm) MoS$_2$ triangular grains merged into a continuous film, using solid source precursors in our system. Larger ~60 μm triangular grains are seen (a) 7 mm from the edge of the substrate, which merge into a continuous film with overlapping grain boundaries at (b) the center of the substrate.

Raman spectroscopy and photoluminescence (PL) are utilized to confirm that 1L MoS$_2$ is present from its characteristic phonon (lattice vibrations) and excitonic (electron-hole pair generation) properties, respectively. In order to ensure strong, well-defined detectable signals, all Raman and PL measurements were performed on 1L MoS$_2$ grown on 90 nm SiO$_2$/Si substrates using a green laser with excitation wavelength of 532 nm. Due to the Lorentzian nature of the Raman (and PL) peaks and the Gaussian nature of the laser spatial intensity, pseudo-Voigt curves were used to capture each peak fit.[49] Figure 1c shows the Raman spectrum of 1L MoS$_2$ grown at 560°C (blue dots). Pseudo-Voigt curves (black lines) fit the Raman peaks of 1L MoS$_2$, namely the in-plane E′ mode at 384.5 cm$^{-1}$ and out-of-plane A$_1$′ mode at 405.3 cm$^{-1}$ with a peak separation $\Delta f$ = 20.8 cm$^{-1}$, which is characteristic of as-grown 1L MoS$_2$[13,50] with slight intrinsic tensile strain due to thermal coefficient of expansion (TCE) mismatch between the 2D material and the underlying SiO$_2$ substrate.[51] Because odd-numbered, few-layer MoS$_2$ (including 1L MoS$_2$) belong to the D$_{3h}$ point group, the two main Raman active modes are denoted E′ and A$_1$′.[13,52] The 2LA(M) peak is seen around 453.3 cm$^{-1}$ and the transverse optical (TO) shoulder peak at 378.8 cm$^{-1}$.[53]

Characteristic photoluminescence (PL) measurements are shown in Figure 1d. We note these probe the optical band gap of monolayer (1L) MoS$_2$, which is smaller than the electronic direct band gap of 1L MoS$_2$ by the exciton binding energy (0.2 to 0.6 eV),[11] and all three of these energies depend on the environmental dielectric screening.[54] Pseudo-Voigt curves (black lines) fit the PL spectrum, identifying the ground state A exciton at 1.847 eV and the charged A$^-$ trion at 1.810 eV. (The A$^-$ trion is a charged exciton with an extra electron coupled to the electron-hole pair.) We did not observe a B exciton in our samples (0.1 to 0.2 eV above the A peak), which has been used to assess non-radiative recombination, with low (or no) B peak suggesting higher sample quality.[55]

To further show that monolayer (1L) MoS$_2$ is grown by chemical vapor deposition (CVD) at 560°C, scanning electron microscopy (SEM) is shown in Figure 1e of the continuous 1L MoS$_2$ film at the center of the substrate with overlapping grain boundaries. Atomic force microscopy (AFM) of 1L MoS$_2$ triangular grains on SiO$_2$/Si, away from the substrate center is also shown in Figure 1f. The MoS$_2$ is measured to be ~0.72 nm thick from the step height, which is in close agreement with the accepted monolayer MoS$_2$ thickness (0.615 nm).[56] Growth residue (most likely caused by excess partially reduced MoO$_3$ precursor) is observed on the MoS$_2$ grains and along the grain edges. Both SEM and AFM indicate that 1L MoS$_2$ is grown without bilayer (2L) MoS$_2$ nucleation.

## ELECTRICAL RESULTS AND DISCUSSION

Transfer length method (TLM) structures[57] were fabricated on 560°C 1L MoS$_2$ as-grown on a 50 nm SiO$_2$ on p$^+$ Si substrate, which also serve as the back-gate for all devices. Figure 2a shows an optical image of the TLM structures as fabricated by electron-beam lithography. The large, square probe pads (20 nm SiO$_2$/2 nm Ti/40 nm Au) lead into the fine contacts (70 nm Au). The colorized scanning electron microscopy (SEM) inset shows the pure Au leads directly contacting the monolayer MoS$_2$ channel, which was etched 18 µm long × 2 µm wide (darker region) after the contact metal deposition. Figure 2b shows the cross-section schematic of the TLM test structure with adjacent contact channel lengths ranging from $L_{ch}$ = 100 nm to 700 nm, forming a series of back-gated field-effect transistors (FETs).

Figure 3a shows the measured drain current $I_D$ (normalized by width) vs. back-gate voltage $V_{GS}$ transfer curves at drain voltages $V_{DS}$ = 0.1 V and 1 V for a 100 nm channel device ($L_{ch}$ = 100 nm). The small arrows mark forward (left $I_D$ branch) and backward (right $I_D$ branch) voltage sweep directions, illustrating small hysteresis and charge trapping between MoS$_2$ and the back-gate SiO$_2$ dielectric. The maximum gate leakage observed was ~1 nA at $V_{GS}$ = 20 V, which is 3-4 orders of magnitude smaller than the drain current in the on-state. A current ratio $I_{on}/I_{off} \sim 10^7$ is observed and the subthreshold slope ($SS$) is estimated $SS \approx$ 1150 mV/dec from the forward sweep at $V_{DS}$ = 0.1 V. We note the $SS$ is relatively high, but can be reduced by reducing the EOT (equivalent oxide thickness) of the gate dielectric below the 50 nm SiO$_2$ used here. Reducing interface charge trap density at the MoS$_2$/dielectric interface[58] by passivating the dielectric[59] or by using a molecular crystal seeding layer[60] can also greatly improve $SS$ when employing cleaner, dedicated fabrication outside general-purpose academic facilities.

Figure 3b shows the measured $I_D$ vs. $V_{DS}$ output curves at gate voltages from $V_{GS}$ = 5 V to 20 V for the same $L_{ch}$ = 100 nm device. The maximum drive current achieved was $I_{D,max} \sim$ 140 µA/µm at $V_{DS}$ = 1 V and $V_{GS}$ = 20 V, which is the highest reported for monolayer MoS$_2$ grown below 600°C. Current saturation is not observed because the threshold voltage $V_T$ is sufficiently negative to keep this device in the linear operating region throughout. Relatively low hysteresis is again observed from the forward and backward voltage sweeps, indicating minimal charge trapping. The measured $I_D$ vs. $V_{GS}$ transfer curves at $V_{DS}$ = 0.1 V for channel lengths from $L_{ch}$ = 100 nm to 700 nm are shown in Figure 3c. The threshold voltage $V_T$ ranges from 2.4 V to 5.8 V, using the linear extrapolation method.[16] The measured $I_D$ vs. $V_{DS}$ output curves at $V_{GS} = 20$ V for channel lengths from $L_{ch} = 100$ nm to 700 nm are shown in Figure 3d. $I_D$ decreases for increasing channel lengths as expected, due to larger channel resistance contribution, for a constant $V_{DS}$ and constant contact resistance ($R_C$).

We next turn to estimating contact resistance ($R_C$) first and then the carrier mobility from TLM structures. In addition to a better estimation of $R_C$, the MoS$_2$ channel sheet resistance ($R_{sh}$) can also be accurately extracted from the slope of the linear TLM extrapolation. The sheet resistance is then used to estimate the effective mobility, $\mu_{eff} = (qnR_{sh})^{-1}$ where $q$ is the elementary charge and $n$ is the carrier concentration. Here $n = C_{ox}(V_{GS} - V_T - V_{DS}/2)/q$ as all our transistors remain in the linear regime, where $C_{ox}$ is the gate dielectric capacitance per unit area (~70 nF/cm$^2$ for our 50 nm SiO$_2$). We note $V_T$ must be individually assessed for each channel in the TLM to account for any device-to-device variation.

Figure 4a shows the linear TLM fit of total measured resistance ($R_{tot} = R_{sh} + 2R_C$, normalized by width) vs. channel length ($L_{ch}$) extracted from Figure 3d at a carrier concentration $n = 6.1 \times 10^{12}$ cm$^{-2}$ (at the same gate overdrive, $V_{OV} = V_{GS} - V_T = 14$ V, not the same gate voltage $V_{GS}$). The $V_T$ is extracted for each channel length in Figure 3c by the linear extrapolation method.[57,61] The $y$-intercept of the linear fit in Figure 4a corresponds to the total contact resistance, $2R_C$. We note that it is important to perform such TLM fits using a wide range of channel lengths (from short, $R_C$-dominated to long, $R_{sh}$-dominated) in order to minimize the $R_C$ extrapolation error. The remaining uncertainty of the linear TLM fit represents the intrinsic device-to-device variation. Figure 4b shows the effective mobility ($\mu_{eff}$) as a function of carrier concentration $n$, obtained from the sheet resistance slope of the TLM plot. At $n = 6.1 \times 10^{12}$ cm$^{-2}$, the effective carrier mobility is $\mu_{eff} = 29 \pm 5$ cm$^2$V$^{-1}$s$^{-1}$, which is comparable to the mobilities reported at higher growth temperatures.[13,16,28–31,33,34] Figure 4c shows the contact resistance ($R_C$, normalized by width) also as a function of carrier concentration $n$. The contact resistance was $R_C = 4.9 \pm 1.3$ kΩ·μm at $n = 6.1 \times 10^{12}$ cm$^{-2}$ with the error bound corresponding to the 95% confidence interval of the line fits.

In Figure 5 we compare the room temperature electron mobility data of our films grown at 560°C with previous reports, for other CVD-grown (and MOCVD-grown[36–41]) 1L MoS$_2$ reported in the literature,[13,16,63–67,18,20,28–30,32,34,62] as a function of growth temperature (up to 900°C). We note some reports are given as effective mobility[13,16,28–30] ($\mu_{eff}$ from TLM, denoted by squares), others are only available as field-effect mobility [18,20,67,32,34,36,62–66] ($\mu_{FE}$, denoted by triangles). The effective mobility values from this work (grown at 560°C for 50-minutes) are marked in yellow. The field-effect mobility values reported from MOCVD growths (at or below 500°C for 8 to 30 hours) are marked in blue.[36–41] For this simple comparison we benchmark two-probe mobility measurements, although more complex four-probe measurements with a threshold voltage correction can yield more accurate mobility values,[68] if current shunt paths through the (invasive) inner voltage probes are avoided.[7] In general, we have found no correlation between carrier mobility and growth temperature, only between growth temperature and MoS$_2$ crystallite size or substrate adhesion. Growth times (at maximum process temperature) are also labeled, indicating significantly longer growths reported for MOCVD to attain full coverage 1L MoS$_2$.

First, comparing the effective 1L MoS$_2$ mobilities ($\mu_{\text{eff}}$, squares), we observe that our values (29 and 33 cm$^2$V$^{-1}$s$^{-1}$ from 560°C growths at a reduced 490 torr pressure) are similar to the $\mu_{\text{eff}}$ from MoS$_2$ grown at higher temperatures, up to 850°C.[13,16,28–30] (Recent simulations have shown mobilities in this range are limited by point defects, most likely charge impurities and sulfur vacancies.[69,70]) Field-effect mobility ($\mu_{\text{FE}}$) data reported for 1L MoS$_2$ grown at 850°C (lighter hollow triangles) are also included as a box-and-whisker plot (average $\mu_{\text{FE}}$ of 34.2 cm$^2$V$^{-1}$s$^{-1}$ outlined in the black box)[16] in good agreement with the 850°C effective mobility. For CVD-grown 1L MoS$_2$ using solid source precursors, our values are the highest reported mobilities to date with a thermal budget below 2-hours at 600°C. One notable difference is that our past 850°C growths have yielded up to ~10% bilayer regions,[16] while such bilayer regions are undiscernible on the 560°C growths presented here (Figure 1). In addition, we have also found the MoS$_2$ grown at 560°C to be more weakly adhered to the SiO$_2$ substrate compared to our previous studies at higher growth temperatures,[13,16,28–30] which necessitated careful electron-beam lithography without exposing the MoS$_2$ samples to water, to avoid delamination during processing.[71,72] We note that adhesion challenges, in addition to the growth residue observed on the 1L MoS$_2$ grains and along the grain edges, could have also led to some of our observed device-to-device variation.

When comparing effective mobility values with field-effect mobilities ($\mu_{\text{FE}}$, triangles) reported across the literature, most $\mu_{\text{FE}}$ reported are lower. This is mainly due to the effects of contact resistance, although material quality, non-ideal contact selection and device fabrication (in academic facilities) could also cause the large $\mu_{\text{FE}}$ variation observed, with no clear trends between studies and across growth temperatures. It is also possible that $\mu_{\text{FE}}$ is overestimated in some studies, depending on the gate-voltage-dependence of the contacts. In other words, a sharp turn-on of back-gated contacts can lead to an overestimated peak transconductance $g_{\text{m}}$ and $\mu_{\text{FE}}$.[68,73,74] Due to these effects of gated contacts, effective mobility ($\mu_{\text{eff}}$) from sheet resistance (TLM) measurements (particularly when reported as a function of carrier density $n$) tend to be more reliable and are preferable rather than $\mu_{FE}$, as the main figure of merit for $MoS_2$.

## CONCLUSIONS

We have demonstrated direct CVD solid source precursor growth of monolayer (1L) $MoS_2$ at 560°C in 50-minutes, with electrical and optical properties very similar to those of CVD $MoS_2$ grown at higher temperatures. Our films are within the 450-to-600°C, 2-hour thermal budget window required for back-end-of-the-line (BEOL) compatibility with modern silicon integrated circuit processing. These new 1L $MoS_2$ films growths were enabled by carefully matching an optimized sulfur flux to sufficient volatile $MoO_{3-x}$ reduced in the presence of the carbon-based PTAS seed layer. Electrical measurements revealed a maximum drive current $I_{D,max} \sim 140$ μA/μm, which is the highest reported for 1L $MoS_2$ grown below 600°C using solid source precursors. The effective electron mobility was extracted with transfer length method (TLM) test structures as $\mu_{eff} = 29 \pm 5$ cm$^2$V$^{-1}$s$^{-1}$ at a carrier concentration of $6.1 \times 10^{12}$ cm$^{-2}$, which is comparable to mobilities reported from films grown at higher temperatures. The results of this work provide a path towards the realization of high quality, thermal-budget-compatible TMDs for heterogeneous integration with silicon manufacturing. These could enable 3D integration of such 2D materials for advanced functionalities in memory or power-gating circuits (with low-leakage BEOL transistors) or high-density logic, in the third dimension.

## METHODS

### 560°C 1L $MoS_2$ Chemical Vapor Deposition Growth

PTAS (perylene-3,4,9,10 tetracarboxylic acid tetrapotassium salt) dissolved in water was deposited as droplets around the perimeter of $1.5 \times 2$ cm chips of thermally grown 50 nm $SiO_2$ on p$^+$ silicon that were initially rinsed with de-ionized (DI) water.[13] The PTAS droplets dry up into "coffee rings" around the perimeter of the substrate before the seeded $SiO_2$/Si chips were placed face-down on top of an alumina ($Al_2O_3$) crucible containing ~2 mg of $MoO_3$ powder (Alfa Aesar, Puratronic 99.9995% purity) spread into a ~1 cm diameter circle. ~100 mg of excess solid sulfur source (Alfa Aesar, Puratronic 99.999% purity) melted in a quartz boat was placed ~26 cm away from the center near the incoming Ar gas flow inlet, all enclosed within a 55 mm inner diameter quartz tube. Before each growth, the system was flushed for 5 minutes using 1500 sccm Ar gas under vacuum before setting the ambient condition to 490 torr at 22 sccm Ar flow rate, adjusting the throttle valve. For the growth temperature cycle, the tube furnace was first ramped from room temperature to 450°C in 10 minutes and then subsequently ramped to 560°C in 5 minutes. The temperature was held at 560°C for 50 minutes before rapidly cooling down to room temperature by opening the furnace hatch. The partially reduced $MoO_3$ source must be cleaned out and replenished after each growth cycle, while the excess solid sulfur source can be re-used in subsequent growths.

**$MoS_2$ Device Fabrication**

Contact probe pads and coarse contacts were defined by e-beam lithography using a bilayer resist stack of 250 nm MMA/50 nm PMMA. This promotes resist undercutting for easier contact metal liftoff which helps maintain $MoS_2$ adhesion to the growth substrate. This was followed by e-beam evaporation of 20 nm $SiO_2$/2 nm Ti/70 nm Au, all layers deposited at 0.5 A/s at a base-pressure of $10^{-7}$ torr. The 20 nm $SiO_2$ layer helps mitigate pad-to-gate leakage through the back-gate dielectric. This was followed by an overnight lift-off in acetone at room temperature. Fine contacts were patterned using e-beam lithography with a 60 nm PMMA 495K/215 nm PMMA 950K bilayer resist-stack which provides fine resolution along with sufficient undercutting for effective liftoff. 70 nm Au was then deposited by e-beam evaporation at a deposition rate of 0.5 A/s at a base pressure below $10^{-7}$ torr without any adhesion layer to achieve a clean contact interface. The liftoff was carried out in Remover PG solvent for 30 minutes at 80°C. $MoS_2$ was patterned into uniform rectangular channels of 2 μm width by photolithography followed by a gentle $O_2$ plasma etch. The $O_2$ plasma etch was carried out at a power of 10 W in a chamber set to 20 mTorr pressure and 20 sccm $O_2$ flow rate for 60 seconds. All electrical measurements were carried out at room temperature in a vacuum probe station at $\sim 10^{-5}$ torr after an in-situ vacuum anneal at 250°C for 2 hours.

**Optical, Electrical, and AFM Characterization**

Raman and PL data were taken using a Horiba Labram with a 532 nm excitation laser. All electrical measurements were performed in the dark and under vacuum ($<10^{-5}$ Torr) using a Keithley 4200-SCS parameter analyzer, in a Janis ST-100 probe station, at room temperature. Scanning electron microscopy (SEM) was conducted using a Thermo Fisher Scientific Apreo S SEM equipped with a NiCol electron column operated in immersion mode. Atomic force microscopy (AFM) was conducted using a Bruker Dimension Icon AFM system operated in tapping mode.

## AUTHOR INFORMATION

**Corresponding Author:** *E-mail: epop@stanford.edu

**Competing Interests:** The authors declare that they have no competing interests.

## ACKNOWLEDGEMENTS

This work was supported in part by ASCENT, one of six centers in JUMP, a Semiconductor Research Corporation (SRC) program sponsored by DARPA; the National Science Foundation (NSF) EFRI 2-DARE Award 1542883; and member companies of the Advanced Materials Enabling Novel Devices (AMEND) focus area of the Stanford SystemX Alliance, an industrial affiliate program at Stanford University. Work was performed in part at the Stanford Nanofabrication Facility (SNF) and the Stanford Nano Shared Facilities (SNSF), which received funding from the National Science Foundation (NSF) as part of National Nanotechnology Coordinated Infrastructure Award ECCS-1542152.

## TABLE OF CONTENTS GRAPHIC

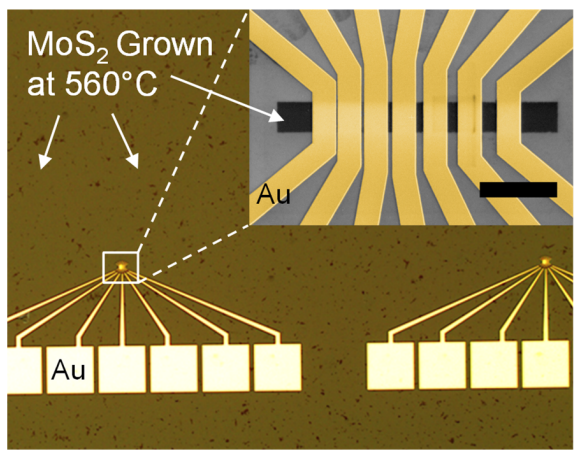
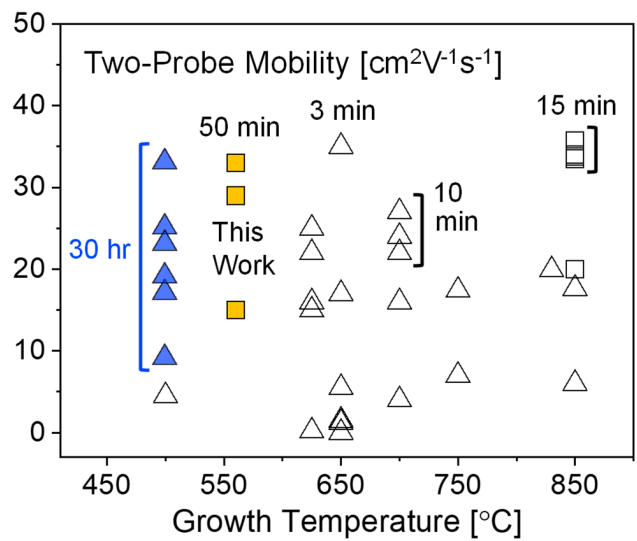

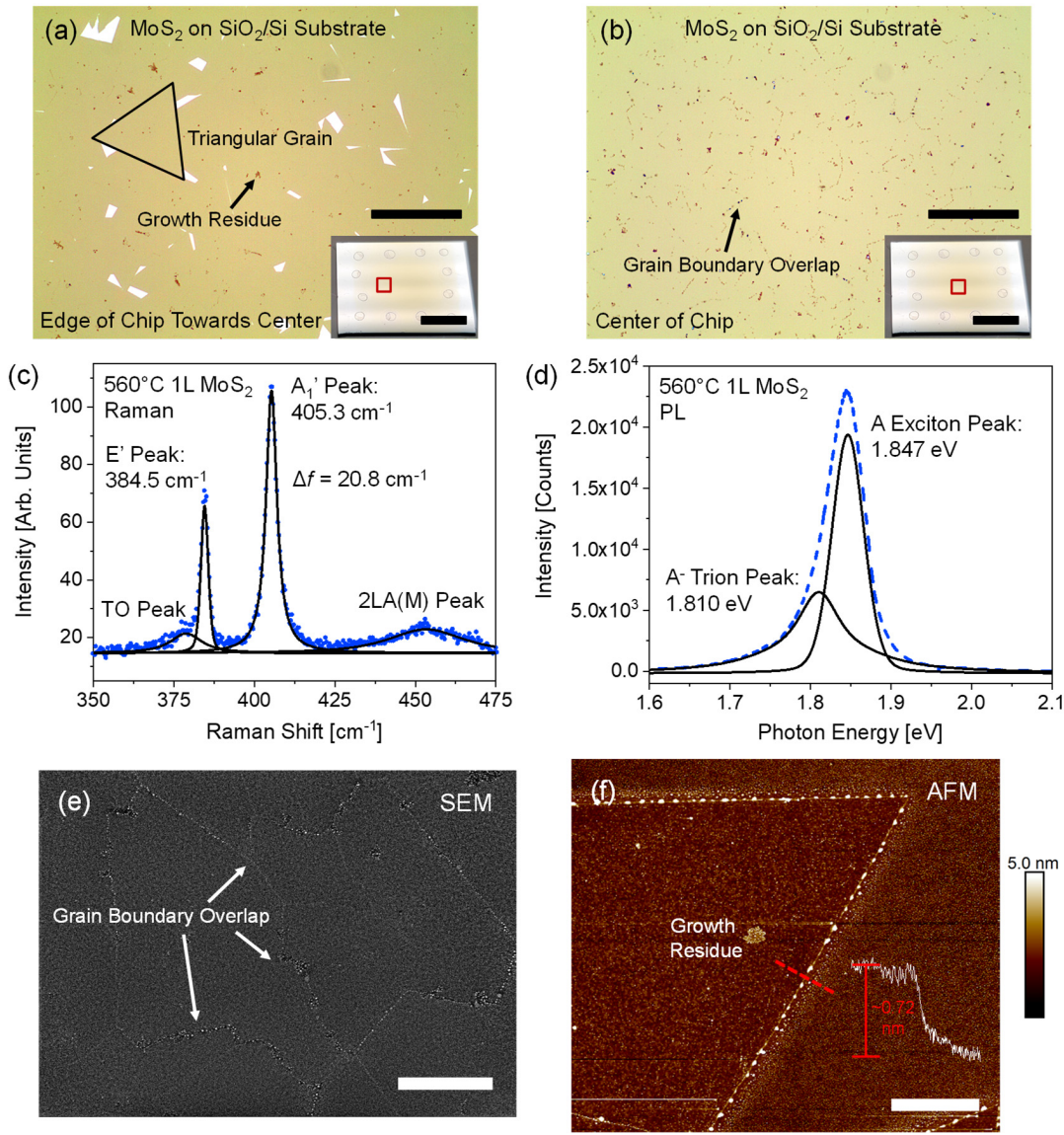

**Figure 1. Monolayer MoS₂ grown at 560°C.** (a, b) Optical images of monolayer (1L) MoS₂ grown by chemical vapor deposition (CVD) at 560°C on SiO₂/Si at (a) 7 mm from edge of the substrate; (b) center of the substrate. Larger ~60 μm triangular grains are seen 7 mm from the edge, which merge into a continuous film with overlapping grain boundaries at the center. Optical image scale bars are 20 μm. Insets show the 1.5 × 2 cm substrates, with red boxes marking the location of the optical images. Circles are "coffee rings" corresponding to the PTAS droplets. Inset scale bars are 7 mm. (c) Raman spectrum of 1L MoS₂ grown at 560°C on SiO₂/Si (blue dots). Pseudo-Voigt curves (black lines) fit the E′ mode at 384.5 cm⁻¹ and the A₁′ mode at 405.3 cm⁻¹ with a peak separation $\Delta f$ = 20.8 cm⁻¹. The longitudinal acoustic 2LA(M) peak is centered at 453.3 cm⁻¹ and the transverse optical (TO) shoulder peak is centered at 378.8 cm⁻¹. (d) Photoluminescence (PL) spectrum of the same 1L MoS₂ (blue dashed line). Pseudo-Voigt curves (black lines) fit the A exciton peak at 1.847 eV and the charged A⁻ trion peak at 1.810 eV. A laser with an excitation wavelength of 532 nm was used for all Raman and PL measurements. (e) Scanning electron

microscopy (SEM) of the continuous 1L $MoS_2$ film at the center of the substrate with overlapping grain boundaries. Scale bar is 3 μm. (f) Atomic force microscopy (AFM) of 1L $MoS_2$ triangular grains on $SiO_2$/Si, away from the substrate center. The $MoS_2$ is measured to be ~0.72 nm thick from the step height, which is in close agreement with the accepted monolayer $MoS_2$ thickness (0.615 nm).[56] Growth residue (most likely caused by excess partially reduced $MoO_3$ precursor) is observed on the $MoS_2$ grains and along the grain edges. Scale bar is 2 μm. Both SEM and AFM indicate that 1L $MoS_2$ is grown without bilayer (2L) $MoS_2$ nucleation.

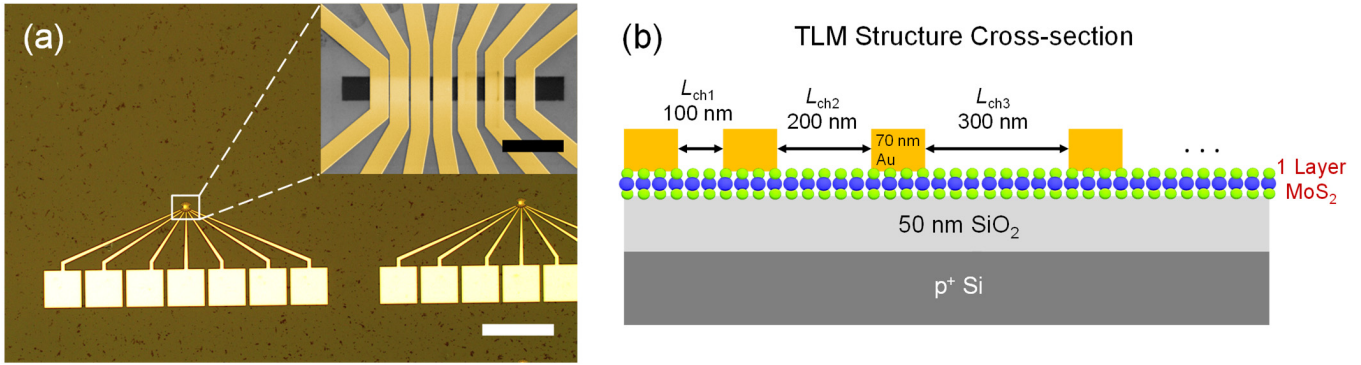

**Figure 2. Transfer length method (TLM) structures.** (a) Optical image of the TLM structures with channels ranging from $L_{ch}$ = 100 nm to 700 nm. The large, square probe pads (20 nm SiO$_2$/2 nm Ti/40 nm Au) lead into the fine contacts (70 nm Au). Scale bar is 100 μm. Inset: Enlarged, colorized scanning electron microscopy (SEM) of the fine leads directly contacting the monolayer MoS$_2$ channel. The channel was etched 18 μm long × 2 μm wide (darker region) after the contact metal deposition. Inset scale bar is 5 μm. (b) Cross-section schematic of the TLM test structure. The gold contact lengths are 1.5 μm wide (not to scale). The Si substrate serves as the gate (G) and pairs of Au contacts as the drain (D) and grounded source (S) in subsequent measurements.

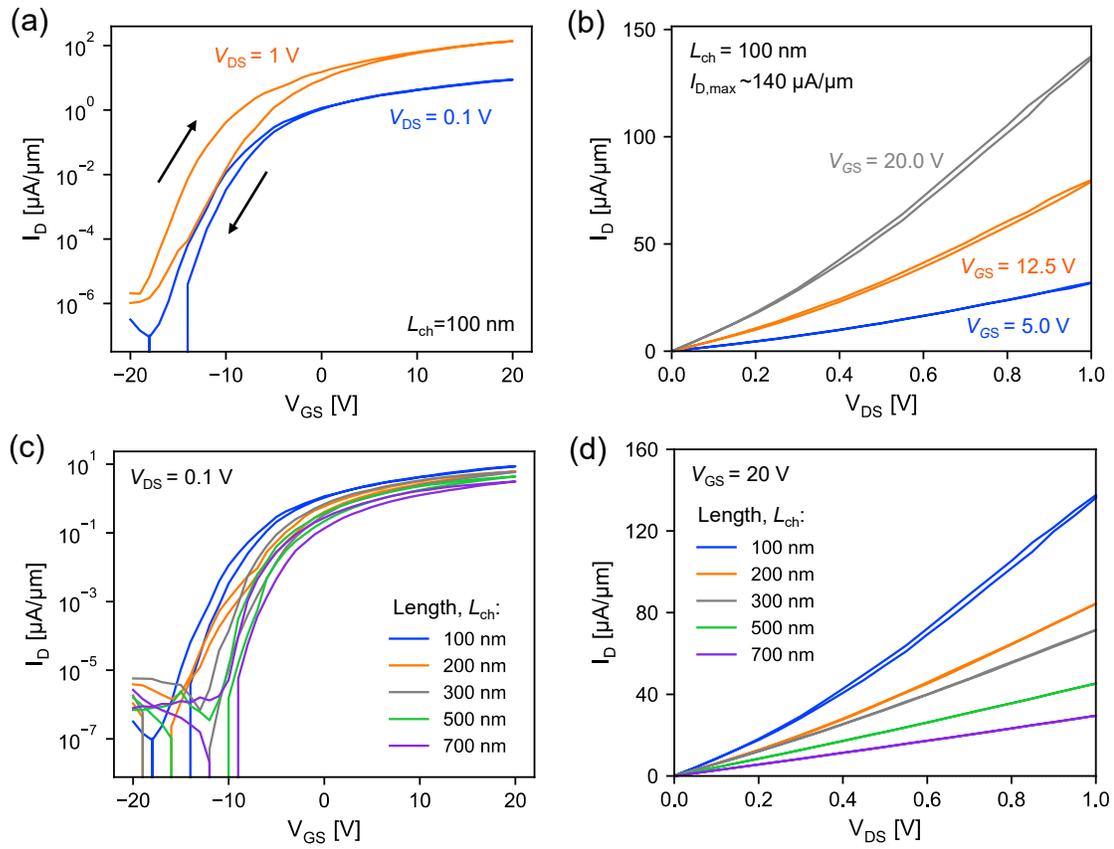

**Figure 3. Electrical characteristics.** Transistors fabricated using monolayer MoS$_2$ grown at 560°C on SiO$_2$/Si substrates. (a) Measured $I_D$ vs. $V_{GS}$ transfer curves at $V_{DS}$ = 0.1 V and 1 V for a 100 nm channel device ($L_{ch}$ = 100 nm). The small arrows mark forward (left $I_D$ branch) and backward (right $I_D$ branch) voltage sweep directions, illustrating small hysteresis. A current ratio $I_{on}/I_{off} \sim 10^7$ is observed and the subthreshold slope is estimated to be $SS \approx 1150$ mV/dec from the forward sweep at $V_{DS}$ = 0.1 V. (b) Measured $I_D$ vs. $V_{DS}$ curves at $V_{GS}$ = 5 V to 20 V for the same $L_{ch}$ = 100 nm channel length device. The maximum drive current achieved was $I_{D,max} \sim 140$ μA/μm at $V_{DS}$ = 1 V. Current saturation was not observed because the threshold voltage $V_T$ is sufficiently negative to keep this device in the linear operating region throughout.[30] (c) Measured $I_D$ vs. $V_{GS}$ transfer curves at $V_{DS}$ = 0.1 V for transistors with channel lengths from $L_{ch}$ = 100 nm to 700 nm. A negative threshold voltage ($V_T < 0$) is observed for all devices in this report. (d) Measured $I_D$ vs. $V_{DS}$ at $V_{GS}$ = 20 V for transistors with channel lengths from $L_{ch}$ = 100 nm to 700 nm. $I_D$ decreases for increasing channel lengths as expected, due to larger channel resistance. All device channels are 2 μm wide and all measurements were performed at room temperature in a vacuum probe station. All electrical measurements shown are double-sweeps (voltage swept up and down), revealing minimal hysteresis.

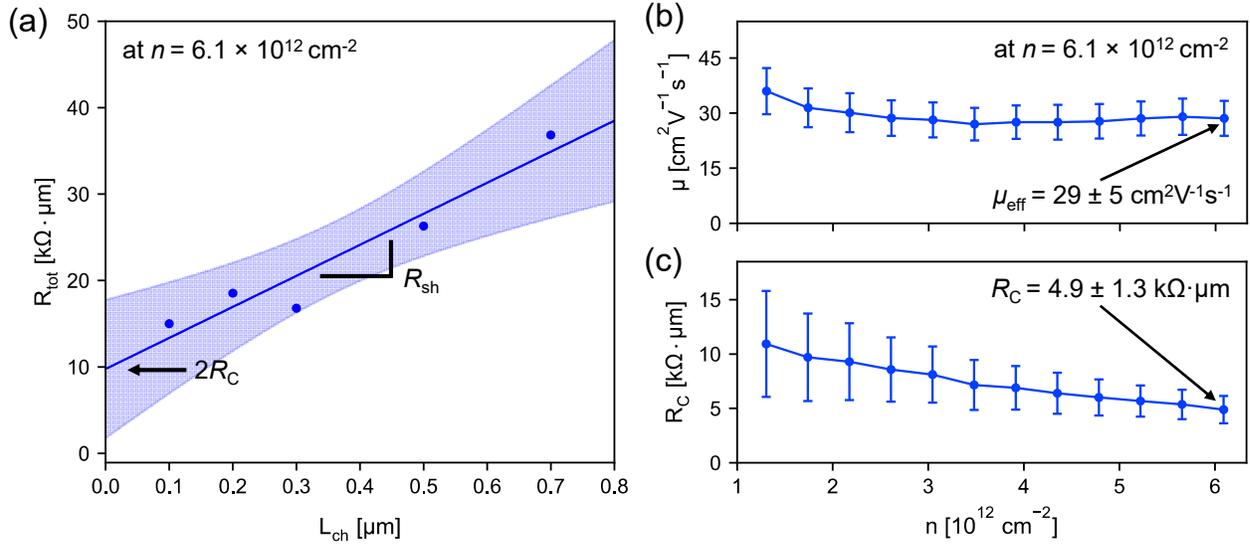

**Figure 4. Sheet resistance and mobility.** (a) Linear TLM fit of the total measured resistance ($R_{tot}$, normalized by width) as a function of channel length ($L_{ch}$) at a carrier concentration $n = 6.1 \times 10^{12}$ cm$^{-2}$, corresponding to the same gate overdrive, $V_{OV} = V_{GS} - V_T = 14$ V. The $y$-intercept corresponds to twice the contact resistance ($2R_C$) while the slope of the line is the sheet resistance ($R_{sh}$). The shaded region marks the 95% confidence bound of the line fit. (b) Effective mobility ($\mu_{eff}$) vs. $n$ as obtained from the sheet resistance. Here, $\mu_{eff} = 29 \pm 5$ cm$^2$V$^{-1}$s$^{-1}$ at $n = 6.1 \times 10^{12}$ cm$^{-2}$, which is comparable to previous reports of monolayer MoS$_2$ grown at higher temperatures (see Figure 5). (c) Contact resistance ($R_C$, normalized by width) as a function of carrier concentration, $n$. The contact resistance was down to $R_C = 4.9 \pm 1.3$ kΩ·µm at $n = 6.1 \times 10^{12}$ cm$^{-2}$. All measurements were taken at $V_{DS} = 0.1$ V and room temperature in a vacuum probe station.

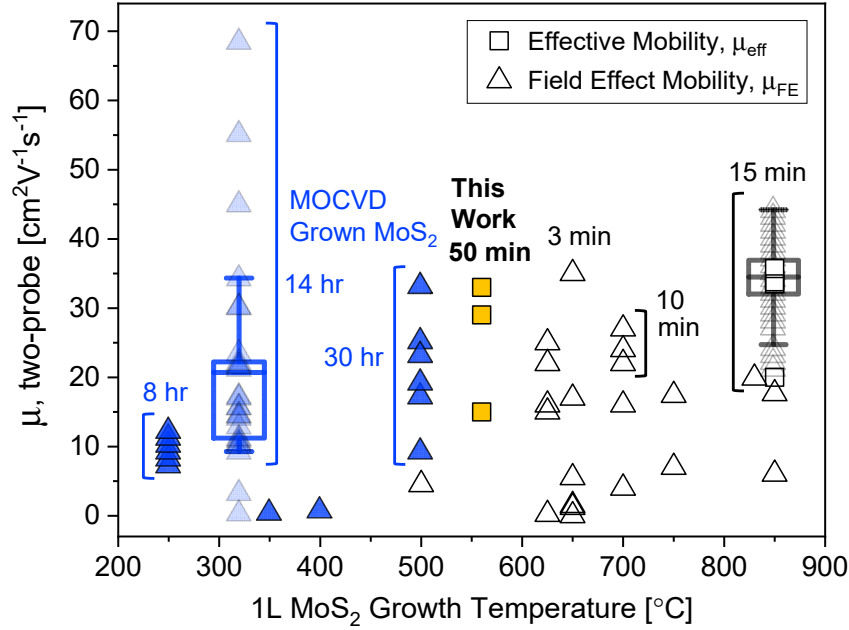

**Figure 5. Monolayer MoS₂ electron mobility vs. growth temperature.** Reported effective mobilities ($\mu_{eff}$ from TLM) are squares and two-probe field-effect mobilities ($\mu_{FE}$) are triangles. The $\mu_{eff}$ from this report are marked in yellow. Metal-organic chemical vapor deposition (MOCVD) data[36–41] are $\mu_{FE}$ marked in blue and solid-source CVD data from the literature[13,16,63–67,18,20,28–30,32,34,62] are hollow symbols. Growth times (at maximum process temperature) are also labeled, indicating significantly longer growths reported for MOCVD to attain full coverage 1L MoS₂. The CVD films in this work were grown at 560°C for 50-minutes, while MOCVD films were grown at or below 500°C for 8 to 30 hours,[36–41] as labeled. The $\mu_{eff}$ from sheet resistance (TLM) measurements tend to be more reliable, while $\mu_{FE}$ could be under- or over-estimated depending on the $V_{GS}$-dependence of gated contacts.[68,73,74] All mobility data reported at room temperature. During review, we became aware of recent MOCVD films grown at 320°C for over 14 hours;[37,38] a summary of $\mu_{FE}$ data for these films (lighter blue triangles) is shown as the box-and-whisker plot at 320°C (average $\mu_{FE}$ of 20.4 cm²V⁻¹s⁻¹ outlined in the blue box).